\documentstyle[epsfig,cite,aps,prl]{revtex}

\twocolumn

\newcommand{\be}{\begin{equation}}
\newcommand{\ee}{\end{equation}}
\newcommand{\bea}{\begin{eqnarray}}
\newcommand{\eea}{\end{eqnarray}}
\newcommand{\lp}{\left(}
\newcommand{\rp}{\right)}
\newcommand{\la}{\left\langle}
\newcommand{\ra}{\right\rangle}
\newcommand{\bm}[1]{\mbox{\boldmath $#1$}}
\newcommand{\bv}{{\mbox{\bm v}}}
\newcommand{\br}{{\mbox{\bm r}}}

\newcommand{\bx}{\mbox{\bm x}}

\renewcommand{\S}{{\cal S}}

\begin{document}

\twocolumn[\hsize\textwidth\columnwidth\hsize\csname
@twocolumnfalse\endcsname

\title{Shear Effects in Non-Homogeneous Turbulence}

\author{F. Toschi$\;^{(1)}$, E. L\'ev\^eque$\;^{(2)}$
and G. Ruiz-Chavarria$\;^{(3)}$}
\address{
${(1)\;}$University of Twente, Department of Applied Physics,\\
P.O. Box 217, 7500 AE Enschede, The Netherlands\\
INFM, Unit\'a di Tor Vergata, Roma, Italy\\
${(2)\;}$Laboratoire de Physique CNRS, ENS de Lyon,
69364 Lyon cedex 07, France \\
${(3)\;}$Departamento de F\'{\i}sica,
Facultad de Ciencias, UNAM, 04510 Mexico D.F., Mexico
}

\date{\today}
\maketitle
\vspace{0.2cm}
\begin{abstract}
Motivated by recent experimental and numerical results,
a simple unifying picture of intermittency in turbulent shear flows
is suggested.
Integral Structure Functions (ISF), taking
into account explicitly the shear intensity, 
are introduced on phenomenological grounds. 
ISF can exhibit a universal scaling behavior, independent of 
the shear intensity.
This picture is in satisfactory agreement with both
experimental and numerical data.
Possible extension to convective turbulence and implication on
closure conditions for Large-Eddy Simulation
of non-homogeneous flows are briefly discussed.
\vskip0.2cm
\end{abstract}

\vskip 0.2cm
]


Statistical properties of turbulent flows are
usually characterized in terms
of the scaling behavior of velocity Structure Functions (SF).
These quantities are
defined as the statistical moments of longitudinal velocity increments
across a separation $\br$ at the location $\bx$:
${D}_p(\bx,\br)=\la   \delta v(\bx,\br) ^p \ra$.
In homogeneous and isotropic turbulence,
${D}_p(\bx,\br)$ only depends on the distance (or scale) $r$.
Experimental and numerical observations support the idea
that ${D}_p(r)$ display universal power-law dependence on $r$
in the so-called inertial range, i.e. ${D}_p(r) \sim r^{\zeta_p}$.
Universality refers here to the scaling exponents $\zeta_p$ being
independent of the stirring process of turbulence.
The $\zeta_p$ values are found to be in disagreement
with Kolmogorov's linear prediction (K41) $\zeta_p=p/3$ \cite{k41}.
The understanding of this correction to K41, usually
referred to as intermittency, has stimulated many phenomenological
and theoretical works during the last 40 years
(see \cite{Frisch} for a recent review).


Only recently, interests in understanding intermittency
in non-homogeneous turbulent flows have started to emerge
(see 
\cite{danaila,Toschi_prl,Toschi_pof,iuso,experimentos,Chilla_etc,Wesfreid,Benzi_shear}).
The major point is to understand how the phenomenology of
intermittency is modified, or can be extended,
in case of non-homogeneous flows.
A common characteristic of such flows, e.g. wall-bounded flows, 
is the presence of a non-zero mean velocity gradient (usually called shear).
Note that shear does not necessarily imply inhomogeneity, e.g. in homogeneous shear,
straining and rotational flows.

Our investigation starts from the following key observations:
i) In presence of a strong shear, intermittency, defined as
the deviation of scaling exponents $\zeta_{p}$
from the linear law, is larger than in homogeneous and
isotropic turbulence.
ii) Relative scaling exponents, measured in very different flows 
but in positions where the shear is strong enough, 
seems to be very similar (universal).

\begin{figure}[!t]
\hskip -.7cm
\epsfig{file=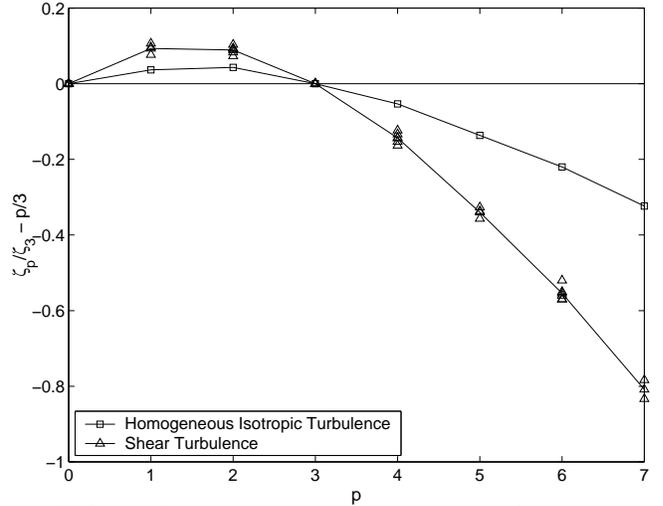,width=\hsize}
\caption{Intermittency corrections to scaling exponents,
$\zeta_p/\zeta_3-p/3$,
in homogeneous and isotropic turbulence ($\Box$),
and several turbulent shear flows ($\triangle$).
Intermittency corrections are significantly larger in presence of
shear and display universality.
}
\label{expo}
\end{figure}

Data in Fig.~\ref{expo} fully confirm our two  key observations. 
They come from very different situations: near the wall
in a channel flow numerical simulation \cite{Toschi_prl,Toschi_pof}
and experiment \cite{iuso},
in the logarithmic sublayer
of a boundary layer flow \cite{experimentos},
near a strong vortex \cite{Chilla_etc}, in the wake of a cylinder \cite{Wesfreid} 
and in a Kolmogorov flow \cite{Benzi_shear}.

In order to provide a theoretical unifying framework
for non-homogeneous turbulent shear flows, we start
from the Navier-Stokes (N-S) equations.
The velocity field can be decomposed into
a mean value (average is meant on time)
plus a fluctuating part:
$\bv(\bx;t)={\overline {\bv}}(\bx)+\bv'(\bx;t)$.
It yields the usual Reynolds decomposition
\be
D_t v'_i+\S_{ij}(\bx) v'_j +
 v_j' \partial_j v'_i - {\overline {v_j' \partial_j v'_i}}
= - \partial_i p' + \nu \Delta v'_i
\label{reynolds}
\ee
with $D_t=\lp \partial_t + {\overline v}_j \partial_j \rp$.
The shear is defined as
$\S_{ij}\lp\bx\rp= \partial_j {\overline v_i\lp\bx\rp}$
and will depend on the mean flow geometry.
In regions where $\S_{ij}=0$, e.g.
very far from the boundaries, turbulence can be considered as homogeneous.
Otherwise, the shear term $\S_{ij}(\bx) v'_j$ must be taken into
account. The results
displayed in Fig.~\ref{expo} show that the presence of this term
modifies significantly the statistical properties of turbulence.
In order to better hilight the physical implication of the shear,
we consider together the second and third terms
of the l.h.s. of  (\ref{reynolds}),
defining the following 
Integral Structure Functions (ISF):
\be
{\tilde D}_p(z,r)= \la \lp  \delta v(z,r)^3
+ \alpha r\cdot \S(z) \cdot \delta v(z,r)^2 \rp^{p/3}\ra,
\label{ISF}
\ee
where $\alpha$ is an empirical prefactor of order one.
These ISF are expected to take into account shear effects,
particularly at large scales (see next paragraph)
and to display a universal behavior.
We consider here {\em generic} situations in which
the shear reduces to $\S(z)=\partial_{z} {\overline v_{x}}$.
Such situation occurs near a rigid wall, where the principal
mean-velocity component aligns in the $x$-direction, 
parallel to the boundary
\cite{Tennekes}.
The shear $\S(z)$ characterizes the variation of $v_{x}$
along the $z$-direction, i.e. as one moves off the wall.
Finally, we consider increments in the direction of the mean flow, i.e.
orthogonal to the shear direction.

ISF reduce on two different SF when either
the first or the second term dominates.
These two terms will exactly balance at scale $L_{\S}$
such that
$ {{\delta v(L_{\S})} / {L_{\S}}} = \alpha~\S.
\label{Ls} $
At scales $r \ll L_{\S}$, shear effects become negligible and 
homogeneous and isotropic scalings are expected. 
Kolmogorov's scaling then yields
$\delta v(r) \sim \epsilon^{1/3} r^{1/3}$, where
$\epsilon$ denotes the mean energy dissipation rate.
By extending this similarity relation to the scale $r=L_{\S}$, one
obtains the usual dimensional estimate for the  
shear length scale $L_{\S} \sim \lp \epsilon/\S^3 \rp^{1/2}$ \cite{Hinze}.
In the logarithmic layer of a plane near-wall flow
it yields $L_\S(z) \sim z$ \cite{experimentos}.
Roughly speaking, $L_{\S}$ can be viewed as the size of small-scale
eddies, whose turn-over time equals the shear time scale $1/\S$,
imposed by the flow geometry and the stirring process
of turbulence at large scales.
Note that this estimate of the shear length-scale stems out
 from dimensional analysis.
In practice, there may be a prefactor in the expression of $L_s$: this
is taken into account by the coefficint $\alpha$.
From previous reasoning it follows that
${\tilde D}_p(r) \sim {D}_p(r)$ for $r\ll L_s$ and 
${\tilde D}_p(r) \sim \lp r \S\rp^{p/3} {D}_{2p/3}(r)$ for $r\gg L_s$.

Another way to see that the central objects, in presence of
shear, are the ISF, comes from the generalization of Yaglom's
equation to homogeneous-shear flows, i.e. with $\S(z)=\S$
\cite{Hinze,Struglia_tesi}:
\bea
\nonumber -{4 \over 5}\epsilon r &=& {D}_3(r) - 6\nu {{d~{D}_2(r)} \over
{dr}} +\\
&-&{2\S\over {r^4}}\int^r_0{dx \; x^4 \;\overline {v_x'(x_0)\cdot v_z'(x_0+x)}}.
\eea
Supposing that this relation can be
generalized, along the same line of idea of the 
Kolmogorov's Refined  Similarity Hypothesis \cite{k62}, one obtains
\be
\delta v^3(r) + \alpha  r \cdot \S \cdot \overline {v_x'(x) \cdot v_z'(x+r)}
\sim \varepsilon(r) \cdot r,
\ee
where $\varepsilon(r)$ denotes the coarse-graining of the energy dissipation field
$\epsilon=\frac{\nu}{2} \sum_{i,j} \lp\partial_i v_j\rp^2$, at
scale $r$.
\begin{figure}[!b]
\hskip -.5cm
\epsfig{file=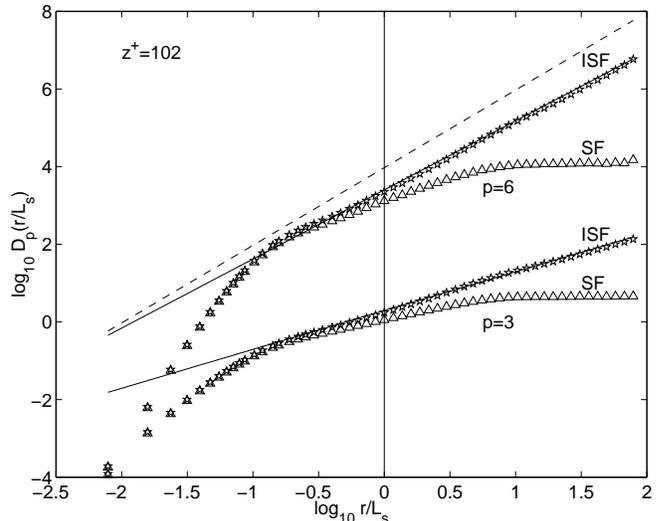,width=\hsize}
\caption{In the turbulent boundary layer at the distance $z^+=102$. 
${D}_3$  and ${D}_6$ ($\triangle$) are compared with ${\tilde D}_3$
and ${\tilde D}_6$ ($\star$).
The scale $r$ has been renormalized by the characteristic shear length-scale $L_s$.
The solid lines passing through ${\tilde D}_3$ and ${\tilde D}_6$
indicate the expected {\it homogeneous and isotropic} power-laws,
respectively $\zeta_3=1$ and $\zeta_6=1.78$.
For comparison, the dashed line has slope $2$.
}
\label{comp1}
\end{figure}
Simplifying, on pure dimensional grounds, the velocity-cross correlation 
$\overline {v_x' v_z'}$, with $\delta v(r)^2$
one ends up again with ${\tilde D}_p(r)$.
Furthermore, we propose the following Refined
Similarity Hypothesis
$$
{\tilde D}_p(r)= \la \lp \delta v(r)^3 + \alpha r \cdot \S \cdot
\delta v(r)^2 \rp^{p/3}
\ra \sim \la \varepsilon(r)^{p/3}\ra r^{p/3}.
$$
\noindent
This formulation is consistent, in the limiting cases of strong and
negligible shear, with some recent findings (see
\cite{Toschi_prl,Toschi_pof}). In addition to that, the ISF should
be able to abridge smoothly between these two limiting regimes, i.e. 
\bea
\label{rksh1}r\ll L_{\S}\;&:&\;\;\;\;{D}_p(r) \sim \la
\varepsilon(r)^{p/3}\ra \cdot r^{p/3}\\
\label{rksh2}{\rm and} \;\;\;\; r\gg L_{\S}\;&:&\;\;\;\;{D}_p(r) \sim \la
\varepsilon(r)^{p/2}\ra.
\eea
Eqn. (\ref{rksh1}) is in agreement with the restoration of
homogeneity and isotropy at small scales. 
For $r \gg L_{\S}$, eqn.  
(\ref{rksh2})
gives ${D}_2(r)\sim const$ (modulo possible logarithmic corrections), yielding
for the energy spectrum $E(k)\sim k^{-1}$ (a relation
suggested long time ago in \cite{Hinze}).

In the previous picture, it is assumed that the shear length scale
$L_\S$ remains larger than the  dissipation length scale $\eta$. 
The dissipation field  
$\varepsilon(r)$ is then expected to display the same scaling properties as
in homogeneous and isotropic turbulence. 
However, in regions where $L_\S \lesssim \eta$ (very
close to the wall) 
it is expected that shear effects act down to the dissipative scale
and therefore  can modify the scaling behavior of $\varepsilon(r)$.
The scalings of ${D}_p(r)$ 
should then also change according to (\ref{rksh2}).   
\begin{figure}[!t]
\hskip -.7cm\epsfig{file=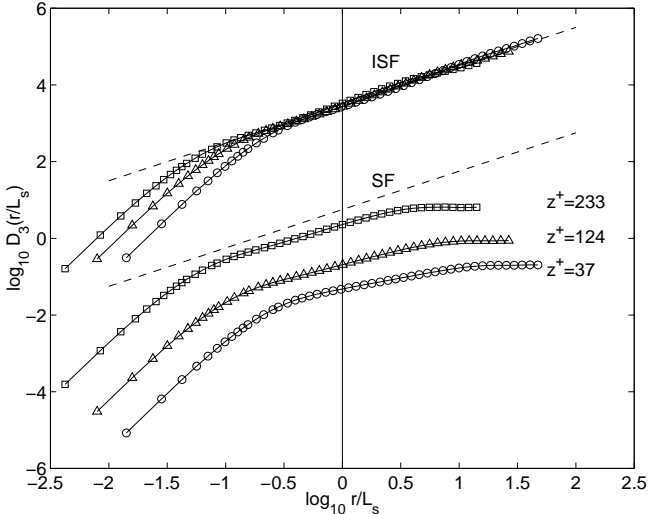,width=\hsize}
\caption{From the boundary layer experiment,
 third-order structure function at various
distances from the wall: $z^+=37$ ($\circ$), $z^+=124$ ($\triangle$)
and $z^+=233$ ($\Box$).
The scaling properties of ${D}_3$ (SF) do depend on the distance $z^+$.
On the contrary, ${\tilde D}_3$ (ISF) displays
the same scaling behavior for all $z^+$.
The dashed line has slope $1$. The curves have been shifted
vertically for convenience.
}
\label{alls3}
\end{figure}
The relevance of Integral Structure Functions for describing the
scaling properties of non-homogeneous shear flows is now tested
on both experimental and numerical data.
The experiment, performed in the recirculating wind tunnel of ENS-Lyon, 
consists in a turbulent boundary-layer flow over a smooth horizontal plate (see \cite{experimentos} 
for details about the experimental apparatus).
Velocity measurements are carried out at various elevations from the
plate in the logarithmic turbulent sublayer \cite{Tennekes}.
Numerical results are obtained from a direct simulation
of the Navier-Stokes equations
in a rectangular channel flow (see \cite{Toschi_prl} for
details).

In Fig.~\ref{comp1}, ${D}_3$ and ${D}_6$,
measured in the logarithmic boundary sublayer, are compared with the
corresponding  ${\tilde D}_3$ and ${\tilde D}_6$.
The shear $\S(z)=\partial_{z} {\overline v_{x}}$
 has been estimated from the mean velocity profile.
In standard non-dimensional variables \cite{Schl}, namely $z^+=v_*z/\nu$ and
$ v^+=\bar v_x/v_*$ where $v_*$ is the characteristic
velocity of the viscous sublayer, our data are well fitted
by the logarithmic profile 
$ v^+(z^+)={(1/{\kappa})}\log(z^+)+B $.
We obtain $\kappa\simeq 0.4$ and $B\simeq 5.26$ 
in agreement with previously 
reported results \cite{experimentos}.
For what concerns the coefficient $\alpha$, all our results have been obtained with the fixed
value $\alpha \approx 0.2$.
Note that the constant $\alpha$ is not (a priori) intended to be universal
but may strongly depend on the geometry and stirring process of the flow.
However, it is expected to remain of order unity. 
In practice, the value of $\alpha$ has been extracted from data 
by requiring that one third-order ISF scale as $r$ (see next paragraph).
Corresponding estimates of $L_s$ are indicated in all figures.
Finally, one must point out that velocity increments have been estimated
in the direction of the mean flow by use 
\begin{figure}[!t]
\hskip -.7cm
\epsfig{file=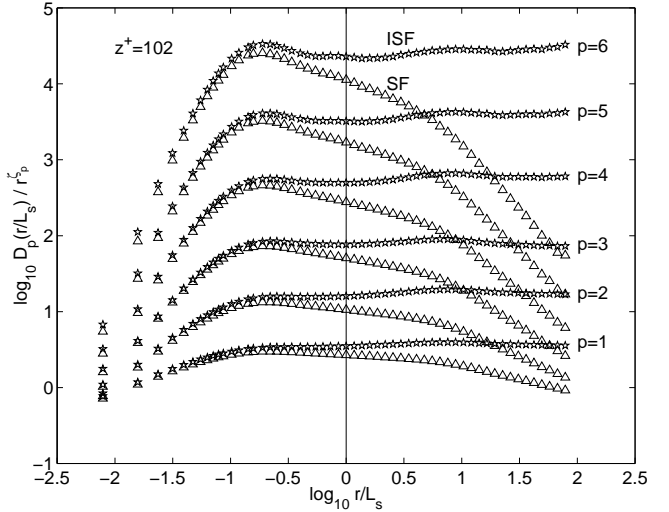,width=\hsize}
\caption{In the experimental boundary layer at $z^+=102$. 
SF and ISF, compensated by homogeneous and isotropic scalings,
are displayed for  p=$1\dots 6$.
}
\label{compensated}
\end{figure}
\noindent
of the Taylor hypothesis.
Both ${D}_3$ and ${D}_6$ exhibit
a power-law dependence on $r$ but
the scaling exponents are clearly different from those observed
in homogeneous and isotropic (h-i)
turbulence, respectively, $\zeta_3 \simeq 1$ and
$\zeta_6 \simeq 1.78$. 
On the other hand, the corresponding ISF exhibit power-laws in good
agreement with h-i scalings (up to very large scales).

In Fig.~\ref{alls3}, third-order SF and ISF are displayed
for various distances from the wall. 
We notice that scaling behavior of ${D}_3$ changes with $z^+$. 
On the contrary, all the corresponding ${\tilde D}_3$ display 
the same power-law scaling with exponent $1$. 
We recall that 
the coefficient $\alpha$ is kept constant and the shear is
estimated from the mean velocity profile; there is 
no adjustable parameter. 

We now report a sharper test: 
SF and ISF, compensated by h-i power-law scalings, 
are displayed in Fig.~\ref{compensated}
for $p=1,\dots,6$ at distance $z^+=102$. 
A departure from h-i scalings is clearly observed for SF.

\begin{figure}[b]
\hskip -.5cm\epsfig{file=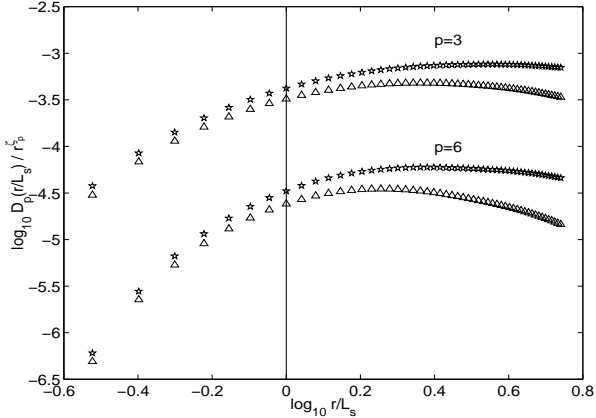,width=\hsize,height=6cm}
\caption{As in Fig.~\ref{compensated} but from numerical data at $z^+=25$. 
}
\label{test_nu}
\end{figure}

On the contrary, ${\tilde D}_p(r)/r^{\zeta_p}$ exhibits a plateau up
to very large scales, indicating that ISF roughly behave as $r^{\zeta_p}$.
In other terms, ISF compensate shear effects
and restore, via the extra term 
$\alpha r\cdot \S \cdot \delta v(r)^2$, the h-i scalings.
Finally, the same test, made on numerical data, 
is reported in Fig.~\ref{test_nu} for the sake of comparison.
Data are obtained at distance $z^+=25$ from the wall, i.e.
where the shear is strong. Results are in reasonable 
agreement with those of Fig.~\ref{compensated}, 
despite the lower resolution of the numerical simulation.
We would like to underline the major points of our study.
Our description relates the scaling properties of velocity fluctuations 
in non-homogeneous shear flows, to those 
of the coarse-grained dissipation rate $\varepsilon(r)$. 
Provided that the shear length-scale remains much larger than the
dissipative length-scale, we are inclined to believe 
that the scaling properties of $\varepsilon(r)$ remain flow-independent; 
the dissipation process, operating on very 
\noindent
small-scales,
is mainly insensitive to the presence of the shear.
We then claim that the scalings of velocity fluctuations 
are different in sheared regions only
because the similarity between $\delta v(r)$ 
and $\varepsilon(r)$ changes form (see (\ref{rksh1}) and (\ref{rksh2})).
When $L_S\lesssim\eta$, shear effects acting down to
dissipative scales are expected to modify the scaling behavior
of $\varepsilon(r)$.  
Nonetheless, we observe in Fig.~\ref{expo} that relative scalings
exponents of velocity structure functions remain universal. 

The introduction of ISF relies on quite simple dimensional arguments. 
However, they have proved to be valuable (preliminary) tools in order
to capture (at leading order) 
shear effects on the scaling behavior of velocity
structure functions. 
We believe that some refinements in the definition 
of ISF may be possible, 
e.g. considering the cross correlation
$\overline {v_x' v_z'}$ instead of $\delta v(r)^2$: however
ISF are of practical interest as they are easily accessible
experimentally. 

A possible extension of ISF applies to thermal turbulent 
convection, the buoyancy term $a g \delta T(r)$ 
playing a similar role of $\S \cdot \delta v(r)$. 
In that case, the ISF would read 
$${\tilde D}^{\mbox{\scriptsize RB}}_p(r)= \la \lp\delta v(r)^3 +
\alpha r \cdot a g \delta T(r) \cdot \delta v(r)\rp^{p/3}\ra$$
and are expected to take into account buoyancy effects at
scales larger than the Bolgiano length-scale.

An important application of our findings concerns Large Eddy Simulations (LES).
Usual eddy viscosity models are known to fail close to the boundary.  
As one moves near the boundary,  
the shear length-scale $L_\S$ becomes smaller and smaller.
While $L_\S$ remains larger than the cut-off scale $\Delta$ of the LES,
usual closure conditions, based on homogeneous 
and isotropic turbulent dynamics, remain acceptable.
However, when $L_\S$ becomes comparable or smaller than $\Delta$,
shear effects must be taken into account and the closure condition
should be modified. 
An alternative consists in decreasing the mesh-size near the
wall so that $L_\S$ always remains larger than $\Delta$. 
More simply, our study suggests to consider
$\lp \delta v(r)^3+ \alpha r \S \delta v(r)^2 \rp^{1/3}$
instead of $\delta v(r)$ in closure relations.
Along this line of idea, 
the Smagorinski's closure condition \cite{smago}
can be generalized in order
to uniformly take into account shear effects. 
The eddy viscosity $\nu_{{\mbox{\footnotesize eddy}}}$ then reads
$$\nu_{{\mbox{\footnotesize eddy}}}= C_s~
\Delta^2\cdot \lp{\overline S}+ \alpha \S \rp,$$
where the extra term $\alpha \S$ takes care of shear effects. 
$C_s$ is the empirical constant 
of the Smagorinski's closure and $\overline S$ denotes
the rate of strain on scale $\Delta$. 

\noindent
{\bf Acknowledgments:}
F. T. would like to thank S. Ciliberto for his kind hospitality at
ENS-Lyon.
G. R.-C. and E. L. acknowledge the ECOS committee and CONACYT
for their financial support under the project No M96-E03.
G. R.-C. also acknowledge DGAPA-UNAM for partial support under the project
IN-107197.

\clearpage
\end{document}